\documentclass[conference]{IEEEtran}
\IEEEoverridecommandlockouts
\usepackage{cite}
\usepackage{amsmath,amssymb,amsfonts}
\usepackage{algorithmic}
\usepackage{graphicx}
\usepackage{textcomp}

\usepackage{xcolor}
\def\BibTeX{{\rm B\kern-.05em{\sc i\kern-.025em b}\kern-.08em
    T\kern-.1660em\lower.7ex\hbox{E}\kern-.125emX}}
\begin{document}

\title{Detecting Conspiracy Theory Against COVID-19 Vaccines\\
}

\author{\IEEEauthorblockN{1\textsuperscript{st} Md Hasibul Amin}
\IEEEauthorblockA{\textit{Department of Computer Science} \\
\textit{University Of Houston}\\
Houston, United States \\
mamin3@uh.edu}
\and

\IEEEauthorblockN{2\textsuperscript{nd} Harika Madanu}
\IEEEauthorblockA{\textit{Department of Computer Science} \\
\textit{University Of Houston}\\
Houston, United States \\
hmadanu@uh.edu}
\and

\IEEEauthorblockN{3\textsuperscript{rd} Sahithi Lavu}
\IEEEauthorblockA{\textit{Department of Computer Science} \\
\textit{University Of Houston}\\
Houston, United States \\
slavu@uh.edu}


\and
\IEEEauthorblockN{4\textsuperscript{th} Hadi Mansourifar}
\IEEEauthorblockA{\textit{Department of Computer Science} \\
\textit{University Of Houston}\\
Houston, United States \\
hmansourifar@uh.edu}

\and
\IEEEauthorblockN{5\textsuperscript{th} Dana Alsagheer }
\IEEEauthorblockA{\textit{Department of Computer Science} \\
\textit{University Of Houston}\\
Houston, United States \\
dralsagh@CougarNet.uh.edu}

\and
\IEEEauthorblockN{6\textsuperscript{th} Weidong Shi}
\IEEEauthorblockA{\textit{Department of Computer Science} \\
\textit{University Of Houston}\\
Houston, United States \\
wshi3@central.uh.edu}

}

\maketitle

\begin{abstract}
Since the beginning of the vaccination trial, social media has been flooded with anti-vaccination comments and conspiracy beliefs. As the day passes, the number of COVID-19 cases increases, and online platforms and a few news portals entertain sharing different conspiracy theories. The most popular conspiracy belief was the link between the 5G network spreading COVID-19 and the Chinese government spreading the virus as a bioweapon, which initially created racial hatred. Although some disbelief has less impact on society, others create massive destruction. For example, the 5G conspiracy led to the burn of the 5G Tower, and belief in the Chinese bioweapon story promoted an attack on the Asian-Americans. Another popular conspiracy belief was that Bill Gates spread this Coronavirus disease (COVID-19) by launching a mass vaccination program to track everyone. This Conspiracy belief creates distrust issues among laypeople and creates vaccine hesitancy. This study aims to discover the conspiracy theory against the vaccine on social platforms. We performed a sentiment analysis on the 598 unique sample comments related to COVID-19 vaccines. We used two different models, BERT and Perspective API, to find out the sentiment and toxicity of the sentence toward the COVID-19 vaccine.
\end{abstract}


\section{Introduction}

Statistics show that COVID-19 is the most rapidly spreading and deadliest virus ever. Since the first COVID-19 case arose in December 2019 in China, 613.4 million people got affected, and more than 6.5 million people have died worldwide, according to WHO as of Sep 2022\cite{world2022covid}. Conspiracy beliefs and misinformation about this make this COVID-19 pandemic worst\cite{ahmed2020covid} \cite{oleksy2021content}. Initially, contradictory statements about wearing the mask and lack of knowledge about the virus between researchers and the government helped to spread misinformation about the COVID-19 virus\cite{montagni2021acceptance}\cite{kowalski2020adherence}. 

Conspiracy theories have a destructive effect on society. It demeans the people’s interest and creates distrust of authentic information \cite{lewandowsky2020conspiracy}\cite{pummerer2022conspiracy}\cite{gao2019target}. Generally, rumors spread faster than original news. For many years, people believed the 9/11 attack was an inside job \cite{lewandowsky2020conspiracy}\cite{douglas2019understanding}\cite{van2018conspiracy}. Global digitization and the sudden social media boom open various doors for users to easily share false and misleading information\cite{shahsavari2020conspiracy}\cite{marcellino2021detecting}. 

Furthermore, due to a lack of information about the effect of the COVID-19 vaccine and general health-related knowledge among people, social media users are promoting false stories about the vaccine’s side effects. A study in Italy found that over 2000 online articles, fake news tended to be shared a million times more\cite{livingston2020md}\cite{douglas2021conspiracy}. On the other hand, few individuals and politicians make this an issue to weaken the ruling party and achieve their personal agenda\cite{oleksy2021content}. 

While researchers said that only the mass vaccination program could overcome this pandemic, this fake and false information creates distrust among the people and leads to vaccine hesitancy \cite{jovanvcevic2020optimism}\cite{montagni2021acceptance}\cite{syed2021survey}. Also, underlying fear about vaccine side effects and instant information about it create vaccine reluctance among the people \cite{hart2018something}\cite{cui2020deterrent}. 

Bill Gates and the 5G conspiracy show the difference in how social media users react and use the conspiracy for different purposes. In South Africa, the 5G controversy attracted minimal interest with debates on the cause and consequence of the virus and did not bother the government \cite{montagni2021acceptance}. At the same time, it was a widespread belief that Bill Gates was behind all of this by launching a vast vaccination program\cite{ahmed2020covid} \cite{gagliardone2021demystifying}. 

Global politicians also promote conspiracy beliefs and misinformation among the people. In addition, this conspiracy belief in the world people leads to distrust towards governments and World Health Organization (WHO), which causes lightness to government regulations about taking vaccines and wearing the mask, and maintaining Physical distance \cite{van2018belief} \cite{bertin2020conspiracy} and thus helps to spread the COVID-19 pervasively. In the beginning, there was also a rumor that COVID-19 was created in LAB and spread by the Chinese government\cite{shahsavari2020conspiracy} \cite{marcellino2021detecting}. 

This study investigates the conspiracy theory against this COVID-19 vaccine by analyzing comments on the online platform. These could reduce the belief in rumors and entertain the vaccination program. We find out the sentiment and toxicity of the different comments using Google Perspective and the BERT model.

\section{RELATED WORK}

Conspiracy theory is quite common in North America. A survey found that about one-quarter to one-third of the population express conspiracy-related opinions\cite{marcellino2021detecting}. In 2020 a survey in the U.S. indicated that about 5\% think that COVID-19 is pre-planned, and 20\% said it could be valid\cite{schaeffer2020look}. Rumors or Conspiracy theories are not always the case of pervasively false beliefs. People sometimes do it intentionally for geopolitical or satisfying individual motives \cite{van2018belief}\cite{bessi2015science}\cite{biddlestone2020cultural}. For example, continuous denial of climate change because of global temperature is gameplay to delay the action against it by demeaning the value of scientific research\cite{lewandowsky2020conspiracy}. 

Conspiracy has its way of supporting and strengthening its views. People start believing in something if they repeatedly see the same news on social media. People get influenced more by social media than by communications\cite{douglas2017psychology}\cite{bessi2015science}. Commercial Branding uses the same strategy to connect people. Conspiracy theories spread rapidly because of the psychological nature of people\cite{biddlestone2020cultural}\cite{ullah2021myths}. Scientific news is less likely to be shared by people than conspiracy pages. Based on the research on 255,225 polarized users of scientific pages, 76.79\% interacted, and among 790,899 conspiracy users, 91.53\% interacted with conspiracy pages in terms of liking’\cite{bessi2015science}.

Besides social media, mainstream T.V. news and political agenda also have a significant influence on Conspiracy beliefs. A report shows that people on T.V. news who are liberal took this pandemic as a national threat\cite{romer2020conspiracy}\cite{biddlestone2020cultural} \cite{hoseini2021globalization}. Opposition for political and legal reasons also plays a vital role in vaccine hesitancy. Initially, few people believed that vaccines did spread from the lab, and it causes autism and infertility in teenage girls, and US Centers for Disease Control and Prevention fought to cover that information\cite{ullah2021myths}. 

Text-Mining is a popular approach to detecting conspiracy theories\cite{ozturk2018sentiment}. Besides, there are numerous machine learning models for text mining, finding out the semantic content or tone of the sentence\cite{marcellino2021detecting}\cite{rieder2021fabrics}\cite{xu2019bert}. On the other hand, Deep neural network performance well in finding out the semantic content of a sentence [21]. BERT and Perspective API also determines the sentence’s sentiment, toxicity, and severity \cite{singh2021sentiment} \cite{zhao2021bert}\cite{hosseini2017deceiving}\cite{devlin2018bert} \cite{sun2019utilizing}. Biddlestone et al., 2020 use Confirmatory Factor analysis (CFA) and Structural equation modeling (SEM) to find out conspiracy belief and intention scale from a set of data collected from different user groups based on age, gender, personality, and education level \cite{biddlestone2020cultural}\cite{elsherief2018peer}. 

Pummerer et al., 2022 randomly survey people to find out who trust and support the government, maintain physical distance, believe in political COVID-19 conspiracy and have a conspiracy mentality\cite{pummerer2022conspiracy}. Gagliardone et al., 2021 performed common word matching in twitter dataset. They find out the most popular keyword, phrases, and hashtags from the Dataset with the help of researchers\cite{gagliardone2021demystifying}. 

Apart from harmful hate speech towards vaccines is also very common on social media. In Nigeria, political parties use these global controversies as a weapon to weaken the interim government \cite{ullah2021myths}\cite{jovanvcevic2020optimism}. On the other hand, the belief that 5G and COVID-19 are linked leads to the burnout of the Cell phone tower in Europe\cite{ahmed2020covid}. Sherief et al. find out hate speech from Twitter data using Perspective API. They find the sentence’s toxicity and attack the commenter by analyzing the score range from 0-1\cite{elsherief2018hate}. Melton et al., 2021 perform lexical sentiment analysis and topics modeling using Latent Dirichlet Allocation on vaccine-related comments on social media\cite{melton2021public}. 

\begin{figure*}[htbp]
\centerline{\includegraphics[width=17cm]{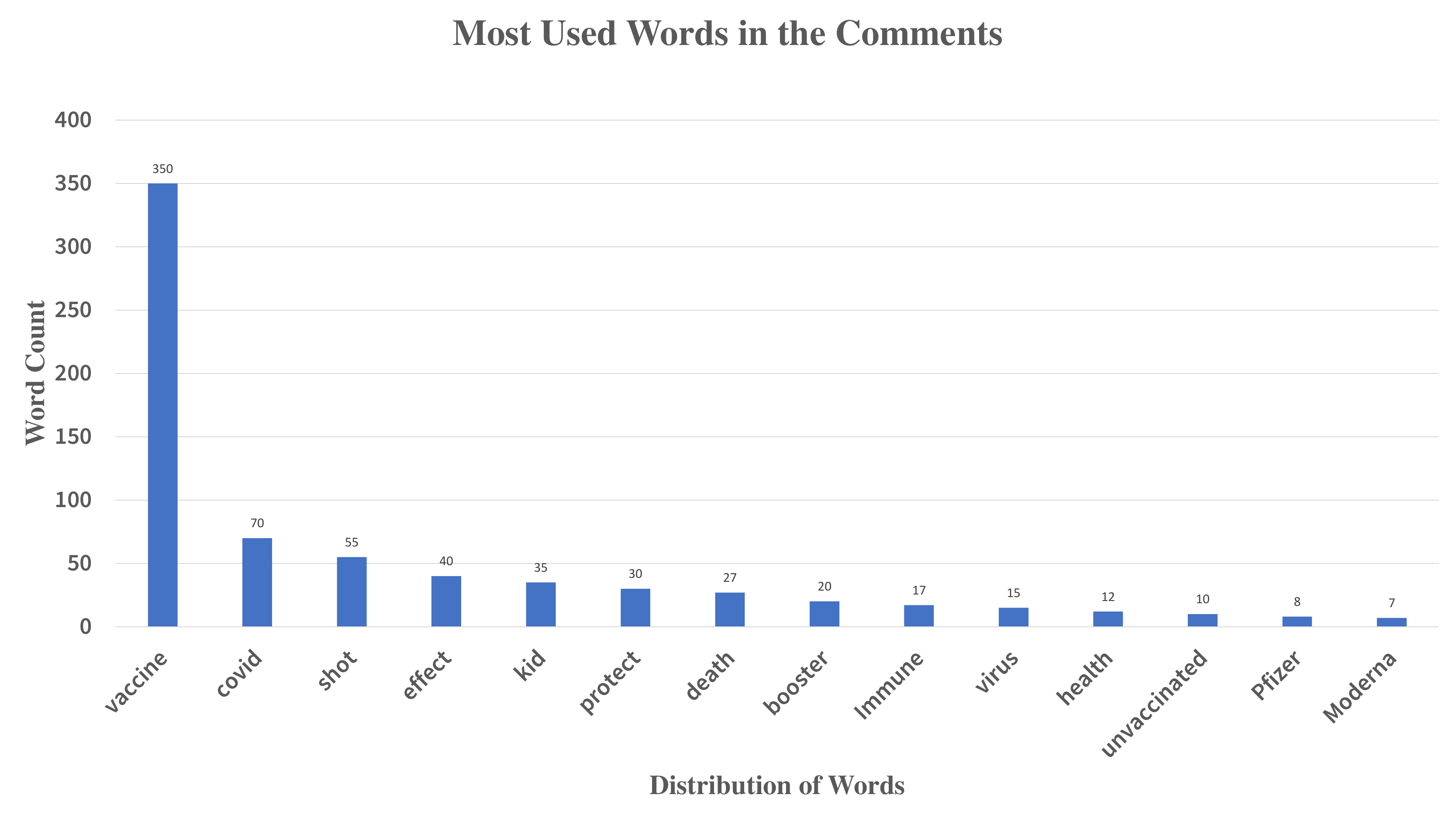}}
\caption{Word Distribution of the Dataset.}
\label{Word Distribution}
\end{figure*}

\section{DATA}
We initially collected 950 users’ comments from various online news portals and their Facebook pages manually. Later, we remove the duplicate comments with an almost similar word. We then manually read and labeled all the comments into two categories, 0 (No) and 1 (Yes). Where 0 means the comment is neutral or in favor of the COVID-19 vaccine, and one means it is against the COVID-19 vaccine. We show some sample data in Table \ref{Sample}. 

Then we clean the data. We removed the comments not in English as our primary focus on English text and finally kept 598 sample comments. We then pre-processed the previously labeled data by removing noise and stop words and converting all the tweets to lower cases. 

This Dataset only includes English comments and is mainly from the North American zone, as our primary focus is to discover conspiracy theories against the COVID-19 vaccine in the United States. All these comments are public and do not include personal information, for example, name, location, or gender. Therefore, it does not go against the privacy act. We have only focused on COVID-19 vaccine-related posts and comments. The data and code is available on GitHub\footnote{https://github.com/AminHasibul/ConspiracyAgaintstCovidVaccines}.

\begin{table}[htbp]
\caption{Sample Data of COVID-19 Vaccines Tweets.}
\begin{center}

\centering
\begin{tabular}{|p{17em}|p{1.3cm}|p{1.1cm}||p{1.1cm}| }
\hline

\textbf{\textit{\hfil Comments}} & \textbf{\textit{Conspiracy }} & \textbf{\textit{\hfil Label}} \\
\hline
After getting vaccine you catch heart diseases & \centering Yes & \centerline{1}   \\
\hline
Fully vaccination can reduce death rate for COVID-19 & \centering No & \hfil 0  \\
\hline
After the second  dose of the Moderna vaccine, people get old & \centering Yes & \hfil 1  \\
\hline
Vaccination can have an impact on gender change & \centering Yes & \hfil  1  \\
\hline
After getting one dose of  the J \& J vaccine to boost the immune system, & \centering No &\hfil 0  \\
\hline
thank you, governor Newsom, for implementing mandates to keep our students and school staff as safe as possible amidst this newest COVID-19 surge.& \centering No &\hfil 0  \\
\hline
\end{tabular}
\label{Sample}
\end{center}
\end{table}

\section{METHODOLOGIES}

Detecting conspiracy theories is a complex problem to solve. Finding conspiracy requires identifying the sentence structure and its sentiment. As the language and structure of the sentences are very discrete in a different language, it is hard to detect and tokenize the word in the correct format \cite{davidson2019racial}. Also, current social media users use emojis, unique jargon, and symbols to express their opinion, which is still a complex task to solve for text classification \cite{ozturk2018sentiment}\cite{zhao2021bert}\cite{rieder2021fabrics}\cite{kovacs2021challenges}.

\begin{figure}[htbp]
\centerline{\includegraphics[width=9cm]{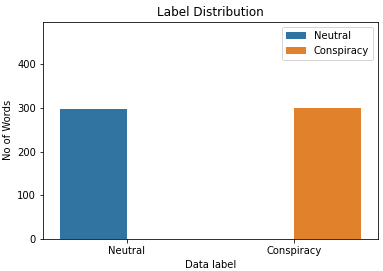}}
\caption{Label Distribution of the Dataset.}
\label{Label Distribution}
\end{figure}

As we had already cleaned and pre-processed our sample in the dataset preparation stage, we first counted the bag of words and word distribution. Then we predict a few incomplete words frequently used in the data and alter them with the proper abbreviation, for example – (vac,vaccn,vcn to the vaccine; CVD, covd to Covid; DCRS, decrees to decrease, fantastic to fantastic; grt to great). We also make sure the data is labeled and distributed evenly. Figure \ref{Word Distribution}  shows the most common word distribution used frequently in tweets or comments. Figure \ref{Label Distribution} shows the label distribution of the data, where we see the same amount of positive and negative comments. 

\begin{table*}[htbp]
\caption{Perspective Scores on Tweets.}
\begin{center}

\begin{tabular}{|p{15em}|p{.8cm}|c|c|p{1cm}|c|c|c|p{1.2cm}|c|}
\hline
\textbf{\textit{\hfil Comments}} & \textbf{\textit{ Label}}
& \textbf{\textit{ Toxicity }}
& \textbf{\textit{ Severe  }}
& \textbf{\textit{ \centering  Identity \\\  Attack}}
& \textbf{\textit{ Insult }}
& \textbf{\textit{\centering  Profanity}}
& \textbf{\textit{Threat }}
& \textbf{\textit{\centering Sexually \\ \hfil Explicit}}
& \textbf{\textit{ Flirtation}} \\
\hline
A vaccine may temporarily affect your period  &\hfil 1    &
0.0561&	0.03674	&\hfil0.04487&	0.09897&	0.04582&	0.07309&\hfil	0.0368&	0.17256\\
\hline
Half of you are crazy. Why would anyone take the vaccine  & \hfil 1 &
0.6758 & 0.35763	& \hfil 0.23627&	0.88076&	0.38093	& 0.15138 & \hfil	0.0479 &	0.26543\\
\hline
Miki Elior Like Colin Powell
died. Fully vaccinated. My point
exactly.&
 \hfil  1 & 0.0458& 0.03464 &\hfil0.02733& 0.03551& 0.01963 &0.25139 &0.0178 &0.2105\\
\hline
I will be happy with my two Pfizer vaccinesA˜ &
 \hfil  0 & 0.0721 & 0.087 &\hfil 0.09076& 0.06088& 0.10754& 0.32974& 0.1811& 0.40821\\
\hline
But your vaccine will protect you! &\hfil 0   &
0.0196&	0.0197&	\hfil0.02333	& 0.05805&	0.0153 &	0.08437&\hfil	0.0163&	0.22616 \\
\hline
The vaccine is a blessing to us  & \hfil 0   &
0.0298&	0.03101&\hfil	0.07103&	0.04766&	0.05292	&\hfil0.09777&\hfil	0.0649	&0.23443\\
\hline
\end{tabular}
\label{Perspective Score}
\end{center}
\end{table*}

After that, we performed sentiment analysis on our labeled data by training the data using two different models, BERT Base and Google Perspective API. In 2018, Devlin et al. proposed Google’s BERT natural language processing model. The BERT model is simple to implement and Powerful. BERT (Bidirectional Encoder Representations from Transformers) is a state-of-the-art machine learning model for NLP tasks\cite{devlin2018bert}. It shows excellent performance in multiple NLP tasks, and pre-training with fine-tuning has become a commonly used method in NLP tasks \cite{pota2020effective}\cite{xu2019bert} \cite{karimi2020improving}. 

BERT model is a multi-layer bidirectional transformer encoder Architecture that completely changed the previous methodologies of pre-training generated word vectors and downstream NLP tasks. The BERT model can take both a single sentence and a pair of sentences as input parameters \cite{devlin2018bert}. It uses Word Piece embedding with a 30000 token vocabulary \cite{hosseini2017deceiving}. BERT trains the sentence-level vector and extracts more information from the context \cite{devlin2018bert}\cite{li2019exploiting}. 

Similarly, Perspective API applies machine learning techniques to identify abusive comments. In addition to the flagship Toxicity attribute, the Perspective model finds the threat, identity attacks, profanity, sexually explicit, and insult. To detect inappropriate comments, Perspective employs machine learning algorithms \cite{hosseini2017deceiving} \cite{rieder2021fabrics}. The algorithms analyze the scores of a phrase based on how influential it is in a conversation. Developers and publishers can use the score to provide feedback to commenters and assist moderators in analyzing comments more quickly\cite{jigsaw2022perspective}.

\section{EXPERIMENT, RESULT \& EVALUATION}

\begin{figure*}[htbp]
\centerline{\includegraphics[width=15cm]{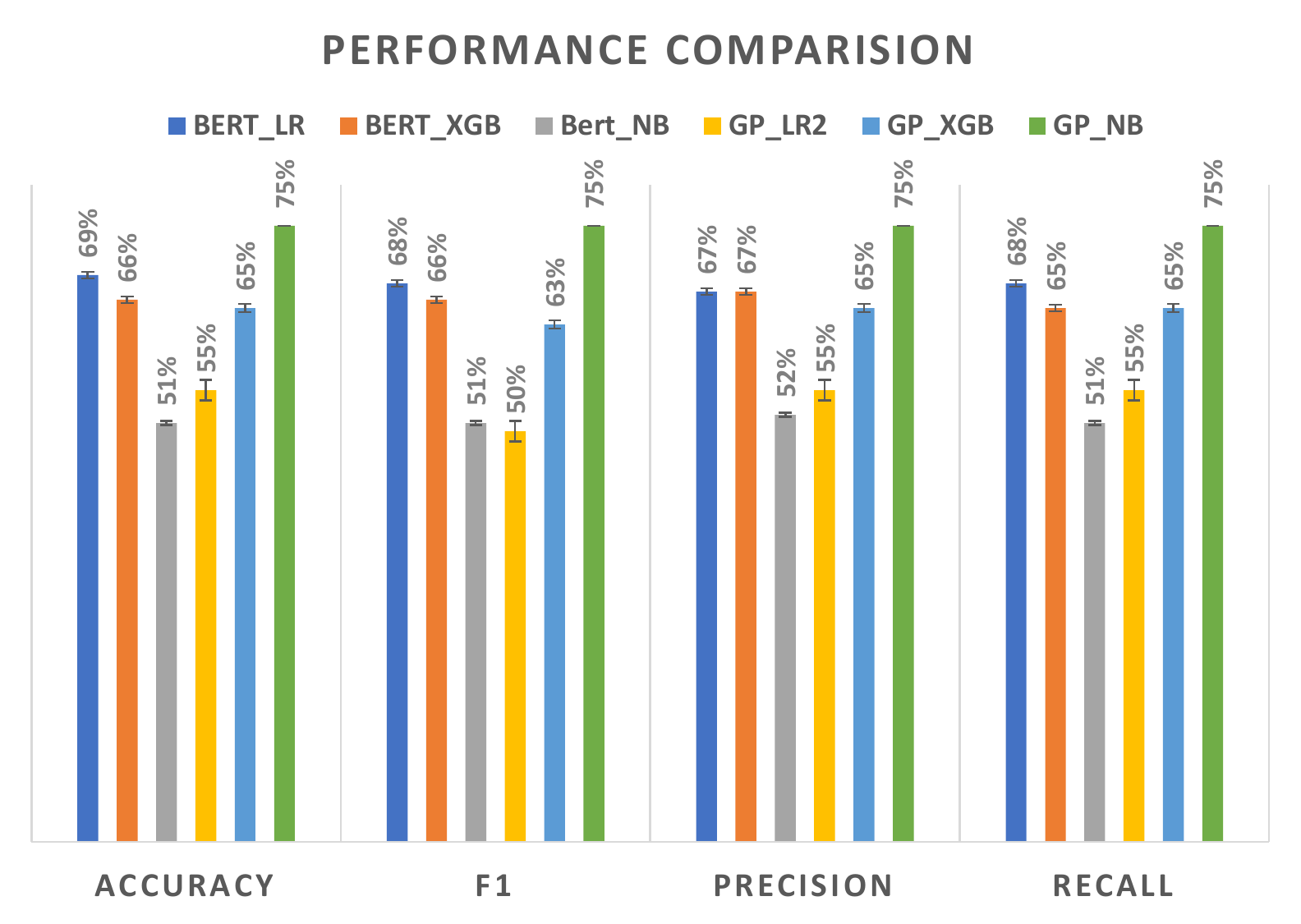}}
\caption{Performance Comparison of Two Models.}
\label{Comparision}
\end{figure*}

We split data into training and validation sets to validate the model performance. We train the model using BERT and Perspective. From Perspective API, we get the different scores (toxicity, severity, threat, attack, sexually explicit, profanity, insult). Table \ref{Perspective Score} shows some samples of Google Perspective score accounting on the comments.

\begin{figure}[htbp]
\centerline{\includegraphics[width=9cm]{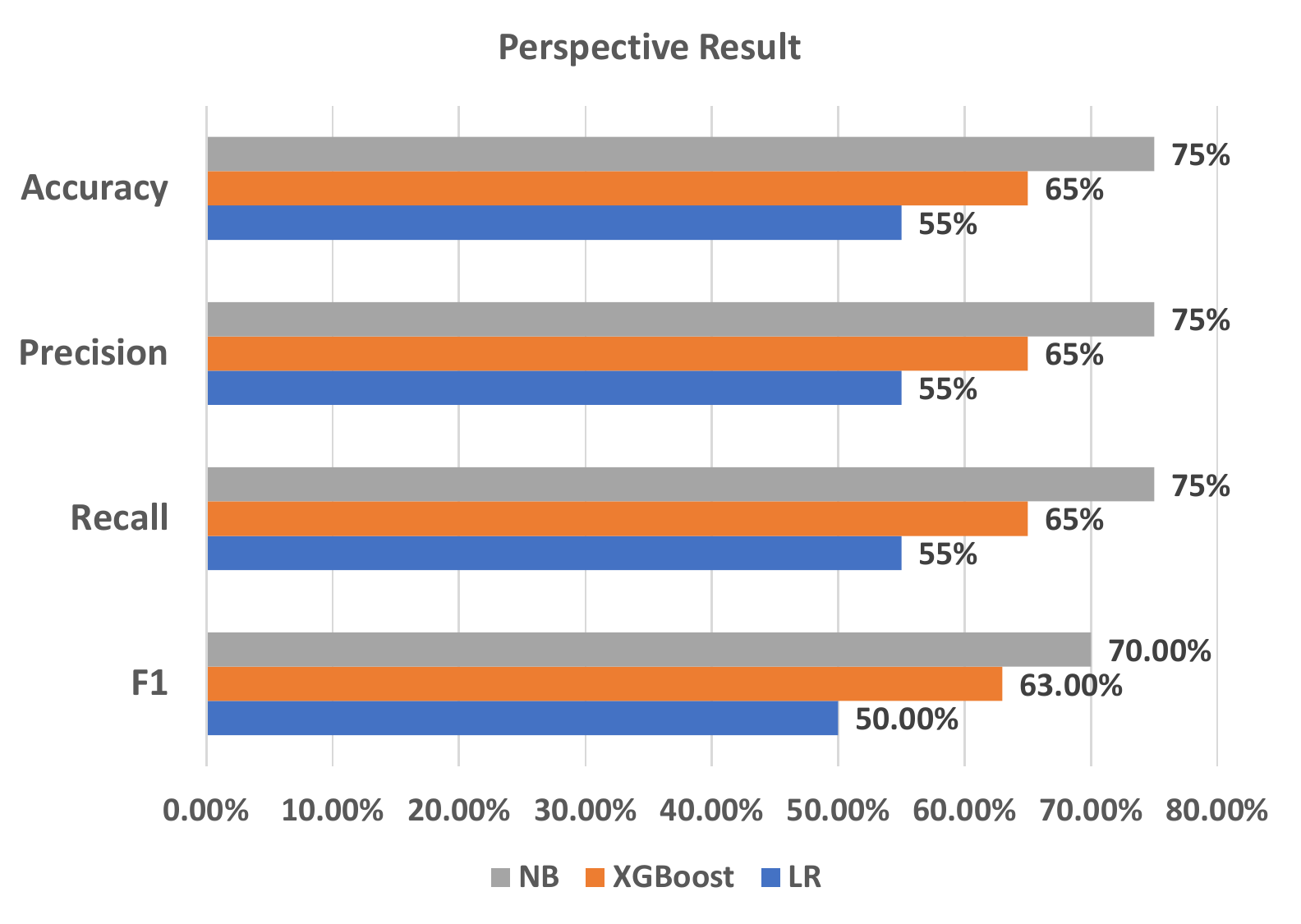}}
\caption{Perspective Accuracy, F1 Score, Recall, and the Precision.}
\label{Perspective}
\end{figure}

\begin{figure}[htbp]
\centerline{\includegraphics[width=9cm]{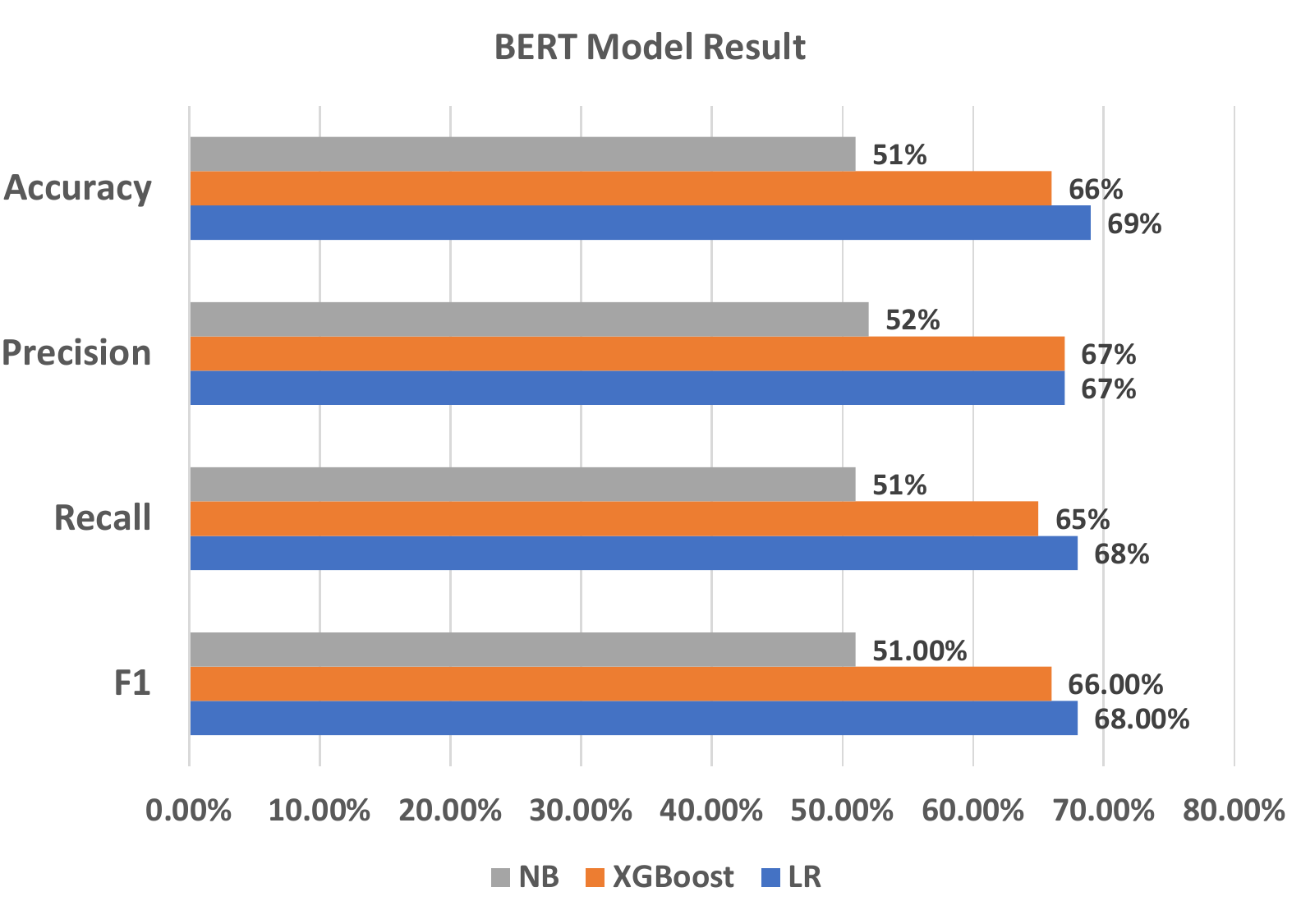}}
\caption{BERT Model Accuracy, F1 Score, Recall, and Precision.}
\label{BERT}
\end{figure}


After that, we performed 10-fold cross-validation on both models. We calculated the Accuracy, F1- score, Precision, and Recall for the newly trained model using Perspective API. Using a Logistic regression classifier, we get an Accuracy of 55\%, a Recall and Precision score of 55\%, and an F1 score of 53\%. For the XGBoost classifier, we get an Accuracy, Recall, and Precision score of 65\%, F1 score of 63\%, and for the Gaussian Naïve Bayes classifier, we get an Accuracy, Recall, and Precision score of 75\%, F1 score of 70\% (Figure \ref{Perspective}).

Then, we train using the BERT model. We used BERT-Base, Uncased, which is a 12-layer model and has 768-hidden layers, 12-heads, and 110M parameters. We then calculated the Accuracy, F1 score, Precision, and Recall for the newly trained model. Using a Logistic regression classifier, we get an Accuracy of 69\%, Recall of 68\%, Precision of 67\%, and F1 score of 68\%. Using an XGBOOST classifier, we get an Accuracy of 66\%, Recall of 65\%, Precision of 67\%, and F1 Score of 66\%. For the Gaussian Naïve Bayes classifier, we get Accuracy, Recall, F1- score of 51\%, and Precision score of 52\% (Figure  \ref{BERT}).

\section{COMPARISON}

Figure \ref{Comparision} shows that Google Perspective with Gaussian Naïve Bayes classifier performance is (75\%) better than others. BERT Model with Logistics Regression gives us an accuracy of 69\%, and XGBoost accuracy is 66\% which is close to L.R. But with Perspective Logistic Regression Performance is very poor, and Naïve Bayes with BERT model performs very poorly. Also, Google Perspective can detect which is abusive and which is not. We also notice an 8-9\% increase in performance for both models by increasing the volume of data in the Dataset. Initially, we performed the same experiment on 400 data, and we got 61\% for BERT Model with the L.R. classifier and 66\% for Perspective with N.B. Classifier. So, we are assuming an increase in performance by adding more data to the Dataset and retraining the model. 

\section{LIMITATION}

We collected our sample data from North American-based users’ comments. The sample data is manually collected and not large enough to perform conspiracy theories for the entire world. So, it does not accurately reflect the thoughts of people from other regions. Also, we did not collect user information, so this dataset is not able to find out the age wise conspiracy belief. On the other hand, it is tough to detect which comments are valid and which are false in some cases. 

\section{CONCLUSION}
Our study analyzed social media’s comments score and public sentiments related to COVID-19 vaccines. As most social media post comments have no structure and many use jargon to express their opinion, it is tough to classify those data \cite{kovacs2021challenges}\cite{al2019detection}. Our study found vaccine hesitancy among people by analyzing public sentiment and comments scores using BERT and Perspective API. Also, we found that comments in favor of vaccines are lower than the comments which are not in favor. Although our study gives an idea about public sentiment in North American regions, further study is needed to detect conspiracy and the user group for the other part of the world to get a more realistic picture of COVID-19 Vaccines.

\section*{Acknowledgment}

The Authors would like to thank the COSC 6376 Cloud Computing Course Fall 21 – University of Houston students for collecting and sharing their data with us. It helps to conduct our research with a more extensive dataset.

\vspace{12pt}

\bibliographystyle{plain}
\bibliography{conference_101719.bib}

\begin{thebibliography}{10}

\bibitem{ahmed2020covid}
Wasim Ahmed, Josep Vidal-Alaball, Joseph Downing, Francesc~L{\'o}pez
  Segu{\'\i}, et~al.
\newblock Covid-19 and the 5g conspiracy theory: social network analysis of
  twitter data.
\newblock {\em Journal of medical internet research}, 22(5):e19458, 2020.

\bibitem{al2019detection}
Areej Al-Hassan and Hmood Al-Dossari.
\newblock Detection of hate speech in social networks: a survey on multilingual
  corpus.
\newblock In {\em 6th International Conference on Computer Science and
  Information Technology}, volume~10, pages 10--5121, 2019.

\bibitem{bertin2020conspiracy}
Paul Bertin, Kenzo Nera, and Sylvain Delouv{\'e}e.
\newblock Conspiracy beliefs, rejection of vaccination, and support for
  hydroxychloroquine: A conceptual replication-extension in the covid-19
  pandemic context.
\newblock {\em Frontiers in psychology}, page 2471, 2020.

\bibitem{bessi2015science}
Alessandro Bessi, Mauro Coletto, George~Alexandru Davidescu, Antonio Scala,
  Guido Caldarelli, and Walter Quattrociocchi.
\newblock Science vs conspiracy: Collective narratives in the age of
  misinformation.
\newblock {\em PloS one}, 10(2):e0118093, 2015.

\bibitem{biddlestone2020cultural}
Mikey Biddlestone, Ricky Green, and Karen~M Douglas.
\newblock Cultural orientation, power, belief in conspiracy theories, and
  intentions to reduce the spread of covid-19.
\newblock {\em British Journal of Social Psychology}, 59(3):663--673, 2020.

\bibitem{cui2020deterrent}
Limeng Cui, Haeseung Seo, Maryam Tabar, Fenglong Ma, Suhang Wang, and Dongwon
  Lee.
\newblock Deterrent: Knowledge guided graph attention network for detecting
  healthcare misinformation.
\newblock In {\em Proceedings of the 26th ACM SIGKDD international conference
  on knowledge discovery \& data mining}, pages 492--502, 2020.

\bibitem{davidson2019racial}
Thomas Davidson, Debasmita Bhattacharya, and Ingmar Weber.
\newblock Racial bias in hate speech and abusive language detection datasets.
\newblock {\em arXiv preprint arXiv:1905.12516}, 2019.

\bibitem{devlin2018bert}
Jacob Devlin, Ming-Wei Chang, Kenton Lee, and Kristina Toutanova.
\newblock Bert: Pre-training of deep bidirectional transformers for language
  understanding.
\newblock {\em arXiv preprint arXiv:1810.04805}, 2018.

\bibitem{douglas2021conspiracy}
Karen~M Douglas.
\newblock Are conspiracy theories harmless?
\newblock {\em The Spanish journal of psychology}, 24, 2021.

\bibitem{douglas2017psychology}
Karen~M Douglas, Robbie~M Sutton, and Aleksandra Cichocka.
\newblock The psychology of conspiracy theories.
\newblock {\em Current directions in psychological science}, 26(6):538--542,
  2017.

\bibitem{douglas2019understanding}
Karen~M Douglas, Joseph~E Uscinski, Robbie~M Sutton, Aleksandra Cichocka,
  Turkay Nefes, Chee~Siang Ang, and Farzin Deravi.
\newblock Understanding conspiracy theories.
\newblock {\em Political Psychology}, 40:3--35, 2019.

\bibitem{elsherief2018hate}
Mai ElSherief, Vivek Kulkarni, Dana Nguyen, William~Yang Wang, and Elizabeth
  Belding.
\newblock Hate lingo: A target-based linguistic analysis of hate speech in
  social media.
\newblock In {\em Proceedings of the International AAAI Conference on Web and
  Social Media}, volume~12, 2018.

\bibitem{elsherief2018peer}
Mai ElSherief, Shirin Nilizadeh, Dana Nguyen, Giovanni Vigna, and Elizabeth
  Belding.
\newblock Peer to peer hate: Hate speech instigators and their targets.
\newblock In {\em Proceedings of the International AAAI Conference on Web and
  Social Media}, volume~12, 2018.

\bibitem{gagliardone2021demystifying}
Iginio Gagliardone, Stephanie Diepeveen, Kyle Findlay, Samuel Olaniran, Matti
  Pohjonen, and Edwin Tallam.
\newblock Demystifying the covid-19 infodemic: Conspiracies, context, and the
  agency of users.
\newblock {\em Social Media+ Society}, 7(3):20563051211044233, 2021.

\bibitem{gao2019target}
Zhengjie Gao, Ao~Feng, Xinyu Song, and Xi~Wu.
\newblock Target-dependent sentiment classification with bert.
\newblock {\em Ieee Access}, 7:154290--154299, 2019.

\bibitem{hart2018something}
Joshua Hart and Molly Graether.
\newblock Something’s going on here: Psychological predictors of belief in
  conspiracy theories.
\newblock {\em Journal of Individual Differences}, 39(4):229, 2018.

\bibitem{hoseini2021globalization}
Mohamad Hoseini, Philipe Melo, Fabricio Benevenuto, Anja Feldmann, and Savvas
  Zannettou.
\newblock On the globalization of the qanon conspiracy theory through telegram.
\newblock {\em arXiv preprint arXiv:2105.13020}, 2021.

\bibitem{hosseini2017deceiving}
Hossein Hosseini, Sreeram Kannan, Baosen Zhang, and Radha Poovendran.
\newblock Deceiving google's perspective api built for detecting toxic
  comments.
\newblock {\em arXiv preprint arXiv:1702.08138}, 2017.

\bibitem{jigsaw2022perspective}
G~Jigsaw.
\newblock Perspective api.
\newblock {\em Acessed May}, 31:2022, 2022.

\bibitem{jovanvcevic2020optimism}
Ana Jovan{\v{c}}evi{\'c} and Neboj{\v{s}}a Mili{\'c}evi{\'c}.
\newblock Optimism-pessimism, conspiracy theories and general trust as factors
  contributing to covid-19 related behavior--a cross-cultural study.
\newblock {\em Personality and individual differences}, 167:110216, 2020.

\bibitem{karimi2020improving}
Akbar Karimi, Leonardo Rossi, and Andrea Prati.
\newblock Improving bert performance for aspect-based sentiment analysis.
\newblock {\em arXiv preprint arXiv:2010.11731}, 2020.

\bibitem{kovacs2021challenges}
Gy{\"o}rgy Kov{\'a}cs, Pedro Alonso, and Rajkumar Saini.
\newblock Challenges of hate speech detection in social media.
\newblock {\em SN Computer Science}, 2(2):1--15, 2021.

\bibitem{kowalski2020adherence}
Joachim Kowalski, Marta Marchlewska, Zuzanna Molenda, Paulina G{\'o}rska, and
  {L}ukasz Gaw{k{e}}da.
\newblock Adherence to safety and self-isolation guidelines, conspiracy and
  paranoia-like beliefs during covid-19 pandemic in poland-associations and
  moderators.
\newblock {\em Psychiatry research}, 294:113540, 2020.

\bibitem{lewandowsky2020conspiracy}
Stephan Lewandowsky and John Cook.
\newblock {\em The conspiracy theory handbook}.
\newblock John Cook, Center for Climate Change Communication, George Mason
  University, 2020.

\bibitem{li2019exploiting}
Xin Li, Lidong Bing, Wenxuan Zhang, and Wai Lam.
\newblock Exploiting bert for end-to-end aspect-based sentiment analysis.
\newblock {\em arXiv preprint arXiv:1910.00883}, 2019.

\bibitem{livingston2020md}
Edward Livingston.
\newblock Md; karen bucher, ma c. coronavirus disease 2019 (covid-19) in italy.
\newblock {\em JAMA-J Am Med Assoc [Internet]}, 2020.

\bibitem{marcellino2021detecting}
William Marcellino.
\newblock Detecting conspiracy theories on social media improving machine
  learning to detect and understand online conspiracy theories.
\newblock Technical report, RAND CORP SANTA MONICA CA, 2021.

\bibitem{melton2021public}
Chad~A Melton, Olufunto~A Olusanya, Nariman Ammar, and Arash Shaban-Nejad.
\newblock Public sentiment analysis and topic modeling regarding covid-19
  vaccines on the reddit social media platform: A call to action for
  strengthening vaccine confidence.
\newblock {\em Journal of Infection and Public Health}, 14(10):1505--1512,
  2021.

\bibitem{montagni2021acceptance}
Ilaria Montagni, Kevin Ouazzani-Touhami, A~Mebarki, N~Texier, S~Sch{\"u}ck,
  Christophe Tzourio, and Confins Group.
\newblock Acceptance of a covid-19 vaccine is associated with ability to detect
  fake news and health literacy.
\newblock {\em Journal of Public Health}, 43(4):695--702, 2021.

\bibitem{oleksy2021content}
Tomasz Oleksy, Anna Wnuk, Dominika Maison, and Agnieszka {\\L}y{\'s}.
\newblock Content matters. different predictors and social consequences of
  general and government-related conspiracy theories on covid-19.
\newblock {\em Personality and individual differences}, 168:110289, 2021.

\bibitem{world2022covid}
World~Health Organization et~al.
\newblock Covid-19 weekly epidemiological update, edition 115, 26 october 2022.
\newblock 2022.

\bibitem{ozturk2018sentiment}
Nazan {\"O}zt{\"u}rk and Serkan Ayvaz.
\newblock Sentiment analysis on twitter: A text mining approach to the syrian
  refugee crisis.
\newblock {\em Telematics and Informatics}, 35(1):136--147, 2018.

\bibitem{pota2020effective}
Marco Pota, Mirko Ventura, Rosario Catelli, and Massimo Esposito.
\newblock An effective bert-based pipeline for twitter sentiment analysis: A
  case study in italian.
\newblock {\em Sensors}, 21(1):133, 2020.

\bibitem{pummerer2022conspiracy}
Lotte Pummerer, Robert B{\"o}hm, Lau Lilleholt, Kevin Winter, Ingo Zettler, and
  Kai Sassenberg.
\newblock Conspiracy theories and their societal effects during the covid-19
  pandemic.
\newblock {\em Social Psychological and Personality Science}, 13(1):49--59,
  2022.

\bibitem{rieder2021fabrics}
Bernhard Rieder and Yarden Skop.
\newblock The fabrics of machine moderation: Studying the technical, normative,
  and organizational structure of perspective api.
\newblock {\em Big Data \& Society}, 8(2):20539517211046181, 2021.

\bibitem{romer2020conspiracy}
D~Romer and KH~Jamieson.
\newblock Conspiracy theories as barriers to controlling the spread of covid-19
  in the us soc sci med. 2020; 263: 113356.
\newblock {\em Search in}, 2020.

\bibitem{schaeffer2020look}
Katherine Schaeffer.
\newblock A look at the americans who believe there is some truth to the
  conspiracy theory that covid-19 was planned.
\newblock {\em Pew Research Center}, 24, 2020.

\bibitem{shahsavari2020conspiracy}
Shadi Shahsavari, Pavan Holur, Tianyi Wang, Timothy~R Tangherlini, and Vwani
  Roychowdhury.
\newblock Conspiracy in the time of corona: Automatic detection of emerging
  covid-19 conspiracy theories in social media and the news.
\newblock {\em Journal of computational social science}, 3(2):279--317, 2020.

\bibitem{singh2021sentiment}
Mrityunjay Singh, Amit~Kumar Jakhar, and Shivam Pandey.
\newblock Sentiment analysis on the impact of coronavirus in social life using
  the bert model.
\newblock {\em Social Network Analysis and Mining}, 11(1):1--11, 2021.

\bibitem{sun2019utilizing}
Chi Sun, Luyao Huang, and Xipeng Qiu.
\newblock Utilizing bert for aspect-based sentiment analysis via constructing
  auxiliary sentence.
\newblock {\em arXiv preprint arXiv:1903.09588}, 2019.

\bibitem{syed2021survey}
SAR Syed~Alwi, E~Rafidah, A~Zurraini, O~Juslina, IB~Brohi, and S~Lukas.
\newblock A survey on covid-19 vaccine acceptance and concern among malaysians.
\newblock {\em BMC Public Health}, 21(1):1--12, 2021.

\bibitem{ullah2021myths}
Irfan Ullah, Kiran~Shafiq Khan, Muhammad~Junaid Tahir, Ali Ahmed, and Harapan
  Harapan.
\newblock Myths and conspiracy theories on vaccines and covid-19: Potential
  effect on global vaccine refusals.
\newblock {\em Vacunas}, 22(2):93--97, 2021.

\bibitem{van2018belief}
Jan-Willem van Prooijen and Karen~M Douglas.
\newblock Belief in conspiracy theories: Basic principles of an emerging
  research domain.
\newblock {\em European journal of social psychology}, 48(7):897--908, 2018.

\bibitem{van2018conspiracy}
Jan-Willem van Prooijen and Mark Van~Vugt.
\newblock Conspiracy theories: Evolved functions and psychological mechanisms.
\newblock {\em Perspectives on psychological science}, 13(6):770--788, 2018.

\bibitem{xu2019bert}
Hu~Xu, Bing Liu, Lei Shu, and Philip~S Yu.
\newblock Bert post-training for review reading comprehension and aspect-based
  sentiment analysis.
\newblock {\em arXiv preprint arXiv:1904.02232}, 2019.

\bibitem{zhao2021bert}
Lingyun Zhao, Lin Li, Xinhao Zheng, and Jianwei Zhang.
\newblock A bert based sentiment analysis and key entity detection approach for
  online financial texts.
\newblock In {\em 2021 IEEE 24th International Conference on Computer Supported
  Cooperative Work in Design (CSCWD)}, pages 1233--1238. IEEE, 2021.

\end{thebibliography}

\end{document}